\documentclass[sigconf]{acmart}
\usepackage{multirow}
\usepackage{times}
\usepackage{soul}
\usepackage{url}
\usepackage{graphicx}
\usepackage{amsmath}
\usepackage{bbm}
\usepackage{booktabs}
\usepackage{algorithm}
\usepackage{algorithmic}
\usepackage{multicol}

\AtBeginDocument{%
  }

\copyrightyear{2022}
\acmYear{2022}
\setcopyright{acmcopyright}\acmConference[CIKM '22]{Proceedings of the 31st ACM
International Conference on Information and Knowledge Management}{October
17--21, 2022}{Atlanta, GA, USA}
\acmBooktitle{Proceedings of the 31st ACM International Conference on Information
and Knowledge Management (CIKM '22), October 17--21, 2022, Atlanta, GA, USA}
\acmPrice{15.00}
\acmDOI{10.1145/3511808.3557341}
\acmISBN{978-1-4503-9236-5/22/10}

\begin{document}

\title{A Gumbel-based Rating Prediction Framework for Imbalanced Recommendation}

 \author{Yuexin Wu}
 \affiliation{
   \institution{University of Memphis}
   \streetaddress{3720 Alumni Ave}
   \city{Memphis}
   \state{TN}
   \country{United States}
   \postcode{38111}
 }
 \email{ywu10@memphis.edu}
 
 \author{Xiaolei Huang}
 \affiliation{
   \institution{University of Memphis}
   \streetaddress{3720 Alumni Ave}
   \city{Memphis}
   \state{TN}
   \country{United States}
   \postcode{38111}
 }
 \email{xiaolei.huang@memphis.edu}



\begin{abstract}
Rating prediction is a core problem in recommender systems to quantify users' preferences towards items. However, rating imbalance naturally roots in real-world user ratings that cause biased predictions and lead to poor performance on tail ratings.
While existing approaches in the rating prediction task deploy weighted cross-entropy to re-weight training samples, such approaches commonly assume a normal distribution, a symmetrical and balanced space. In contrast to the normal assumption, we propose a novel Gumbel-based Variational Network framework (GVN) to model rating imbalance and augment feature representations by the Gumbel distributions.
We propose a Gumbel-based variational encoder to transform features into non-normal vector space.
Second, we deploy a multi-scale convolutional fusion network to integrate comprehensive views of users and items from the rating matrix and user reviews. 
Third, we adopt a skip connection module to personalize final rating predictions. 
We conduct extensive experiments on five datasets with both errors- and ranking-based metrics.
Experiments on ranking and regression evaluation tasks prove that the GVN can effectively achieve state-of-the-art performance across the datasets and reduce the biased predictions of tail ratings.
We compare with various distributions (e.g., normal and Poisson) and demonstrate the effectiveness of Gumbel-based methods on class-imbalance modeling.
The code is available at \href{https://github.com/woqingdoua/Gumbel-recommendation-for-imbalanced-data}{\color{blue}{https://github.com/woqingdoua/Gumbel-recommendation-for-imbalanced-data}}.
\end{abstract}



\maketitle
\section{Introduction}
Rating prediction, an essential task in recommender systems, measures how much a user likes an item.
The task is to find the best products or services according to user interests or preferences by user reviews.
Rating- and review-based models are the main methods in the rating prediction task.
The rating-based models capture the latent feature of users and items based on the rating matrix and then predict the ratings.
The review-based models capture semantic information of user preferences based on reviews, 
enriching the latent features and alleviating the sparsity problem of the rating matrix.

\begin{figure}
\centering
\includegraphics[width=0.35\textwidth]{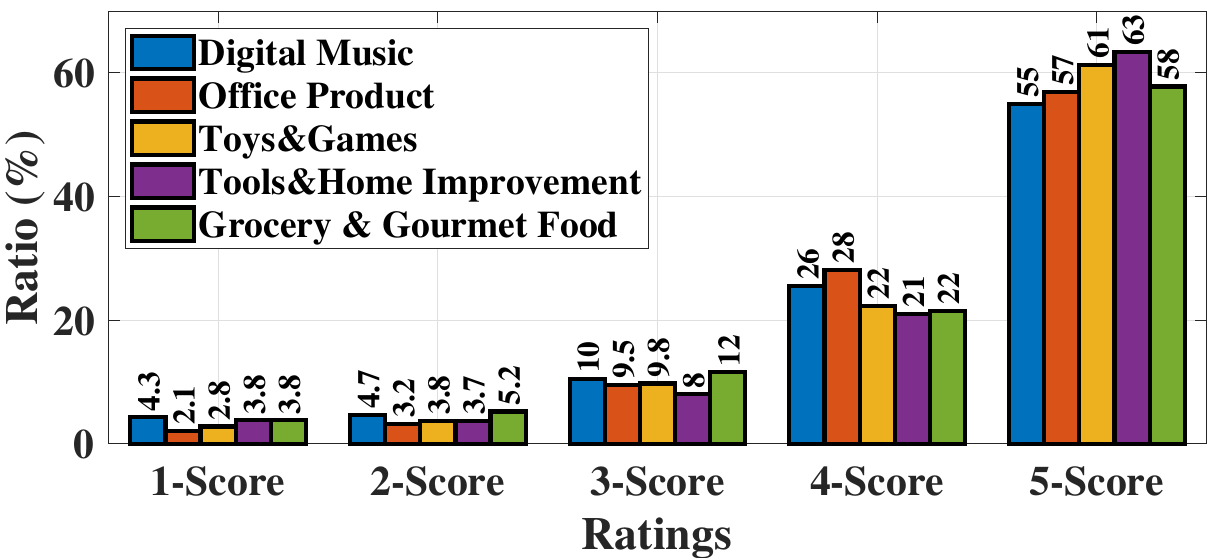}
\caption{The rating distributions of the Amazon reviews.}
\label{fig:samplex}
\end{figure}

However, the imbalance distribution of ratings naturally exists in recommendation datasets and can significantly reduce the effectiveness of recommendation models.
Long-tail distribution is the main reason that causes imbalanced ratings. 
In Figure~\ref{fig:samplex}, we illustrate the issue by five categories of the Amazon review data,\footnote{\url{http://jmcauley.ucsd.edu/data/amazon/}} a common-used dataset for rating prediction evaluations~\cite{he2016ups}.
All subcategories of the review data illustrate imbalanced distributions that tail classes (i.e., 1-, 2-, and 3-score) are much smaller than head classes (i.e., 4- and 5-score).
Studies~\cite{zhao2020improving, wei2021towards, mahadevan2021a} in recommendation systems have shown the imbalanced distribution can lead to misclassification and reduce model effectiveness on the minority rating classes.
While the ways to balance data primarily focuses on re-weighting classes, existing studies~\cite{yang2021adversarial} of imbalance class in the rating prediction primarily hold the assumption of normal distribution, which may not capture skewed feature representations of imbalance ratings.

\begin{table*}[!ht]
    \centering
	    \begin{tabular}{c || c c c c c | c c c c c}
		\multirow{2}*{Dataset} &|U| &|I|&|R| & density(\%) & |T|  & \multicolumn{5}{c}{Ratings\%}\\
		&&&&&& 1 & 2 & 3 & 4 & 5 \\
		\hline\hline
		Musical Instrument &1,429 &900 &10,261 &0.798 &91.66 &1.97&2.31 &7.29 &20.37 &68.06\\
		Toys Game &19,412 &11,924 &167,597 &0.079 &101.92 &2.91&3.80 &9.47 &21.66 &62.16\\
		Digital Music &5,541 &3,568 &64,706 &0.327 &199.91 &4.33 &4.71 &10.04 &26.03 &55.01\\
		Video Game &24,303 &10,672 &231,779 &0.089 &208.66 &2.43 &2.91 &8.26 &23.37 &63.03\\
		Yelp  &1,631 &1,633 &78,966 &2.960  &189.31 &3.84 &8.22 &19.18 &41.69 &27.43\\
        \end{tabular}
    \caption{Statistical summaries of datasets: |U|, number of users; |I|, number of items; |R|, review counts; density, $\frac{|I|*|U|}{|R|}$; |T|, average number of tokens per review. We present percentage distributions of user ratings.}
\label{tb:data}
\end{table*}

\begin{figure*}[ht!]
\centering
\includegraphics[width=0.8\textwidth]{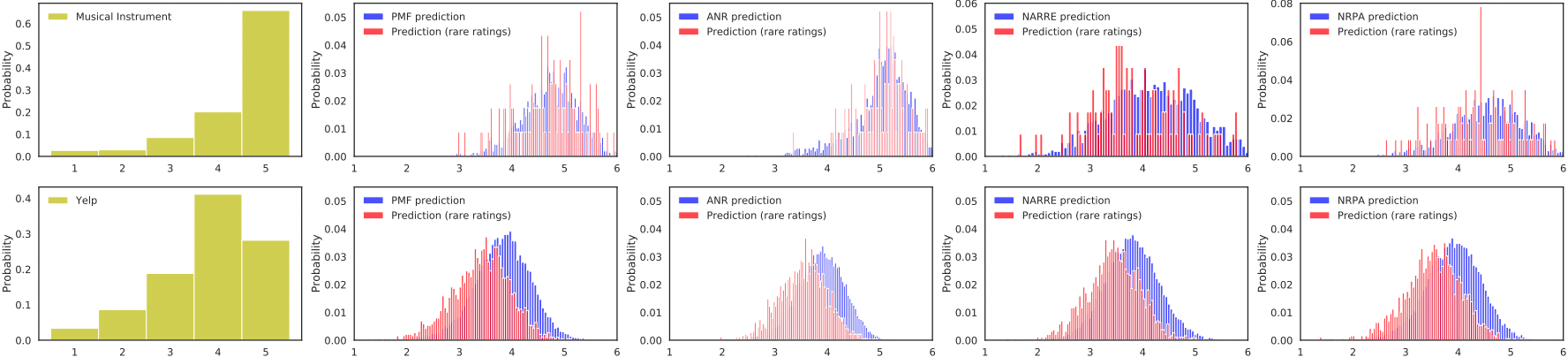}
\caption{The score distributions compare ground truths (yellow), prediction on overall (blue)  and rare ratings (red) across four baselines. Y-axis represents the ratio of predicted rating scores, and the X-axis indicates the rating range.}
\label{tb:sample}
\end{figure*}

In this study, we propose a Gumbel-based Variational Network framework (\textit{GVN}) to model the imbalanced feature distribution of ratings.
Gumbel distribution is a type of extreme value distributions for modeling probabilities of minimums or maximums that are rare events in imbalanced datasets~\cite{gumbel1935valeurs}.
However, there is no prior study has utilized the approach to solve the imbalanced class challenge in the rating prediction task.
The network employs Gumbel distribution and variational neural networks to model tail classes of ratings.
In GVN, we utilize neural networks to learn the Gumbel distribution parameters $\alpha$ and $\beta$.
Then we characterize users and item features by sampling some values from the learned Gumbel distribution. 
Such a method can accurately discover the hidden probability patterns on the tail and head classes.
While the Gumbel-based network models tail classes of user ratings, it does not explicitly consider user preferences and item properties.
To achieve this, we propose a multi-scale convolutional fusion layer to summarize user preferences and item properties from the reviews.
The layer utilizes filters with different sizes aiming to learn diverse semantic information from user reviews.
Moreover, to prevent the gradient vanishing in the Gumbel-based network, we deploy a skip connection to combine the layer outputs with a user rating distribution. 
The skip connection allows us to capture multiple views of user preferences and personalize rating prediction classifiers.
We compare our approach with baselines on standard review data, Amazon reviews~\cite{he2016ups}.
The evaluation results on an online review data show that our proposed approach improves rating predictions on tail classes of ratings and achieves the best performance overall. 
We summarize our contributions as follows:
\begin{itemize}
    \item We propose a Gumbel-based variational network to model imbalance ratings by transforming features into the non-normal vector space. The experiments show that our method outperforms all the compared baselines on both error-based and ranking-based metrics. To our best knowledge, this work is the first attempt to solve the biased prediction problem in recommendation systems using Gumbel distribution.
    \item We apply our proposed framework on tail rating classes and demonstrate the effectiveness of our method in promoting model performance on the tail ratings. Our experiments have demonstrated that the proposed framework can generally promote existing methods on the tail classes, which indicates the Gumbel-based vector space fits better for imbalance rating modeling.
    \item We conduct extensive comparisons on various distributional vector spaces (e.g., Gaussian and Poisson). Extensive experiments prove the effectiveness of Gumbel distribution for the imbalanced data challenge. Our ablation analysis proves that Gumbel-based approaches have more advantages in modeling tail classes in rating prediction tasks.
\end{itemize}

\section{Data}
We retrieve four Amazon review data~\cite{he2016ups} (Musical Instrument, Digital Music, Toys Game, Video Game) and Yelp review~\cite{yelpinc} in this study.
Each data contains four primary information: user, item, ratings, and review documents.
The Amazon data provides users with more than 5 reviews, we keep the same filtering for the Yelp for consistency.
We tokenize the review documents by the NLTK~\cite{bird2004nltk} and summarize the detailed statistics in Table~\ref{tb:data}.
The table shows that the datasets have varied numbers that provide ideal cases to validate the robustness of different models.
We can find that the ratings have an imbalance distribution that positive ratings count the majority.
The imbalance ratings motivate us to examine how the imbalance impact predictions of various rating prediction models.

To examine impacts of imbalance ratings, we conduct an exploratory analysis on the four existing models (PMF~\cite{mnih2008probabilistic}, ANR~\cite{chin2018anr}, NARRE~\cite{chen2018neural}, NRPA~\cite{liu2019nrpa}) that achieved state-of-the-art performance.
The analysis follows the standard evaluations of the previous studies that split each dataset into training set (80\%), validation set (10\%), and test set (10\%) randomly.
We train models on the training set, tune hyperparameters on the validation set, and evaluate models on the test set.
We report distributions of ground truths and the four models' predictions over the Amazon and Yelp data in Figure~\ref{tb:sample}.

We can generally observe that normal distributions of models' overall and rare rating predictions indicating an overfitting towards the majority ratings.
Comparing to the skewed distributions of imbalance ratings, the predicted distributions fail to align with ground truths.
However, the recommender systems~\cite{mnih2008probabilistic, chin2018anr, chen2018neural, liu2019nrpa} for the rating prediction generally hold the normality assumptions, which may not be effective for modeling the imbalance ratings.
The observation motivates us to propose a method that models class imbalance from a non-normal distribution perspective, the Gumbel-based framework.

\section{GVN Framework}

\begin{figure*}[ht]
\centering
\includegraphics[width=0.75\textwidth]{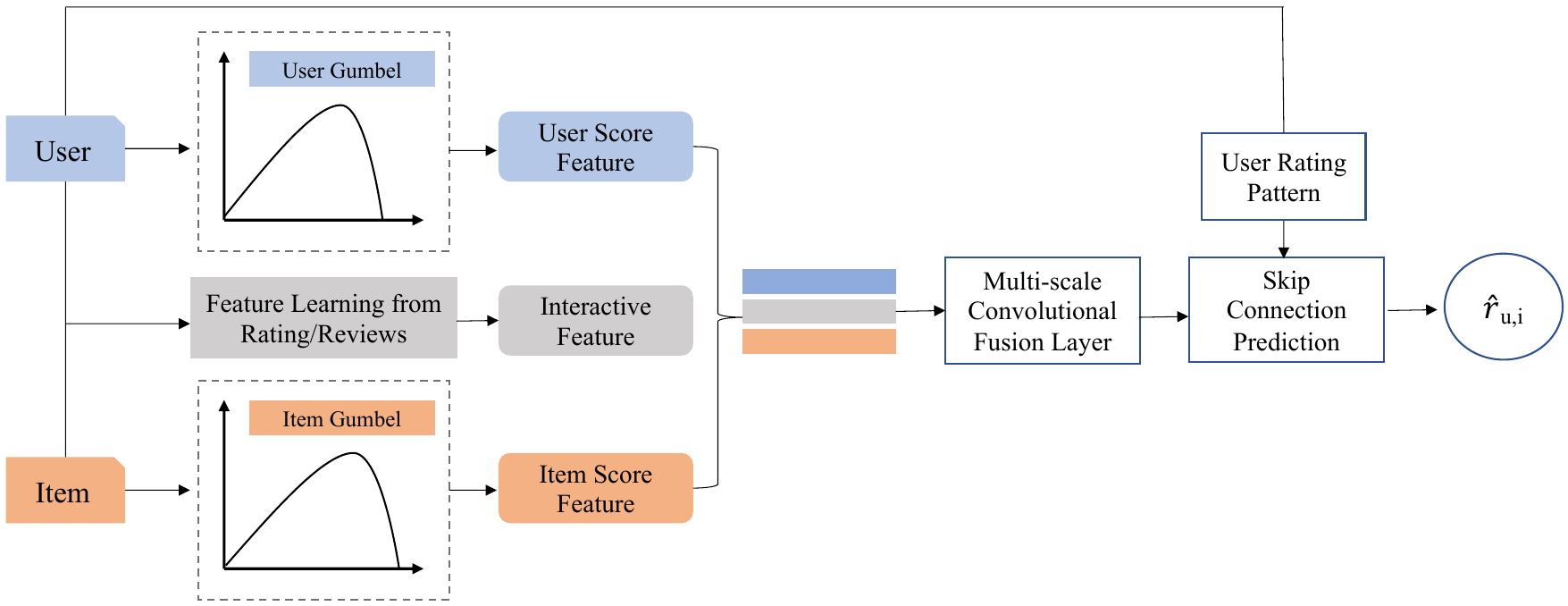}
\caption{Overview of Gumbel-based Variational Network framework.}
\label{fig:overview}
\end{figure*}

We present our Gumbel-based Variational Network framework (\textit{GVN}) in Figure~\ref{fig:overview}.
Our approach contains three major components: 1) a Gumbel-based variational network to learn score features from both items and users;
2) a feature fusion module that uses the multi-scale convolutional fusion layer to combine score features and the interactive features between users and items as the final representation; 
3) a skip connection rating prediction module to personalize the rating scores by comprehensive views of user preferences and item characteristics.
While existing methods~\cite{jang2017categorical} utilizes Gumbel-Softmax for predictions, our work builds a variational encoder network with Gumbel function to transform feature representations into non-normal vector space.
In addition, there is no prior study has transformed features into non-normal space via Gumbel distribution.

\subsection{Feature Encoding}

We obtain three types of features to the model, user, item, and reviews.
For the user and item features, we take rating frequency of users and items as our input.
Given all the historical ratings of a user $u_i$, denoted as $\mathbf{r_{u, i}} = \{r_{1},r_{2}, ..., r_{n}\}$, where $n$ is the number of ratings for the user $u_i$, we denote user rating frequencies of the user as $\mathbf{freq^i_u} \in \mathbb{R}^{1 \times c}$ as follows:
 \begin{equation}
     \mathbf{freq^i_u} = \{\frac{n_{1}}{n}, \frac{n_{2}}{n}, ..., \frac{n_c}{n}\} \ 
 \end{equation}
, where $\{n_1,n_2, ..., n_{c}\}$ is the number of each categorical rating in $\mathbf{r_{u, i}}$, and $c$ is the number of scoring criteria (e.g., $c=5$ denotes that there are $5$ optional ratings for each item). 
Similarly, we can follow the same step to obtain the rating frequencies of the item $j$, denoted as $\mathbf{freq_{i}^{j}}$.
Given user ($i$) and item ($j$), rating-based methods predict the score ($r_{i,j}$) with the frequencies ($\mathbf{freq^i_u}$ and $\mathbf{freq_{i}^{j}}$), while the review-based models will take reviews of the item, $\mathcal{D}_{u}^{i}=\left\{d_{u, 1}^{i}, d_{u, 2}^{i}, \ldots, d_{u, k}^{i}\right\}$, where $k$ is the number of reviews by a user $u$. 
We encode reviews by feature encoders and enrich the review features by both users' and items' $\mathcal{D}_{u}^{i}$ and $\mathcal{D}_{i}^{j}$ to predict a rating score, $r_{i,j}$.
The feature encoders enable the flexibility of our framework to select different base models to extract feature representations.



\subsection{Gumbel-based variational network}
\begin{figure}[ht]
\centering
\includegraphics[width=0.30\textwidth]{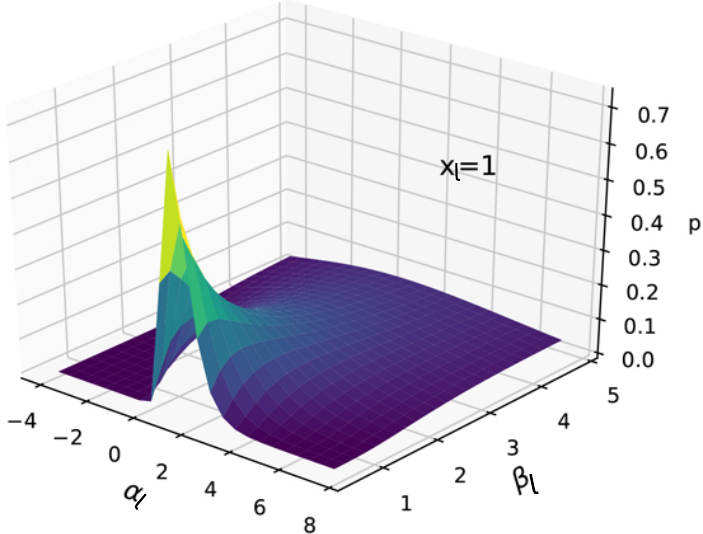}
\caption{Gumbel distributions with $\alpha,\beta$ and fixed $x=1$}
\label{fig:gumbel_pdf}
\end{figure}

Inspired by the VAE network~\cite{kingma2014auto}, we propose a novel Gumbel-based variational network based on Gumbel probability density function (PDF) to capture feature distribution patterns and transform features into non-normal space.
We input feature representations (e.g., $\mathbf{freq_{u}^{i}}$ and $\mathbf{freq_{i}^{j}}$) from the previous step into parallel fully connected feed-forward networks, obtaining the $\mathbf{\alpha}$ and $\mathbf{\beta}$.
The Gumbel PDF generates $\mathbf{p_{u}^{i}} \in \mathbb{R}^{1 \times c}$ to represent rating patterns by the following equation: 
 \begin{equation} 
    p_{u,l}^{i} = \frac{1}{\beta_{l}} \exp \left[\frac{x_{l}-\alpha_{l}}{\beta_{l}}-\exp
     \left(\frac{x_{l}-\alpha_{l}}{\beta_{l}}\right)\right], \beta_{l} > 0.
 \end{equation}
, where the $p_{u,l}^{i}$ represents a hidden possibility of the $l$-th rating of $\mathbf{p_{u}^{i}}$, $l$ is the number of ratings, $x_{l}$ is a hyperparameter ($x_{l} \in \mathbf{x} = [1,2,3,4,5]$) responding to rating categories.
We smooth the $\beta$ as $\beta_{l} = \beta_{l}^{2} + \bigtriangleup \delta$, where $\bigtriangleup \delta$ is a small threshold.  
Empirically, we set $\bigtriangleup \delta$ as $0.5$ in our experiments.
Similarly, we can follow the same step to obtain rating distribution patterns of the item $j$, denoted as $\mathbf{p_{i}^{j}}$.

The Gumbel PDF function can smooth the skewed rating distributions by optimizing $\alpha$ and $\beta$ vectors. 
Scale ($\beta$) and location ($\alpha$) are two parameters of the probability density function (PDF), as shown in Figure~\ref{fig:gumbel_pdf}, which can balance probabilities across different rating scores.
The location controls the function shape, and the scale controls the graph density.
This has recently found success in newly structured prediction tasks~\cite{jang2017categorical}.
The Gumbel mechanism promotes the rare samples' hidden probability giving a higher $\alpha_{l}$ in Figure~\ref{fig:gumbel_pdf}.
Thus, the Gumbel variational network can capture a probability even though a user never gives a lower rating by smoothing the rare ratings.
Note that our proposed Gumbel PDF function is different from the available softmax by jointly balancing rare and frequent rating distributions.\footnote{\url{https://pytorch.org/docs/stable/generated/torch.nn.functional.gumbel_softmax.html}}
The value of $\alpha_{l}$ decides whether it locates in a activation district and the value of $\beta_{l}$ decides the extent of activation.
A large $\beta_{l}$ has smaller probability when given a fixed $\alpha_{l}$.

\subsection{Feature Fusion Module}
We design a feature fusion layer to incorporate the representations from various views of interactive features (denoted as $\mathbf{s_{u,i}}$) with the Gumbel-based score features ($\mathbf{p_{u}^{i}}$ and $\mathbf{p_{i}^{j}}$).
The module aims to capture comprehensive views of interactive features from rating-based (or reviewed-based) models and the score features from Gumbel-based score feature learning.
In this work, we introduce a multi-scale CNN-based fusion layer to achieve the goal. 
This method can extract the feature information from the multiple input vectors and effectively learn the hidden correlation across different input vectors. 

We utilize a based model (rating-based or reviewed-based) to model a user-item interactive feature $\mathbf{s_{u,i}} \in \mathcal{R}^{e \times 1}$.
$e$ is the length of interactive feature, deciding by the based model.
We project the feature $\mathbf{s_{u,i}}$ into the same feature dimension as $\mathbf{p_{u}^{i}}$ by linear function and obtained $\mathbf{t_{i,j}}$.
Then, we stack $\mathbf{p_{u}^{i}}$, $\mathbf{p_{i}^{j}}$ and $\mathbf{t_{i,j}}$ into the same feature map, denoted as $\mathbf{z_{i,j}} \in \mathcal{R}^{c \times 3}$.
Our multi-scale convolutional fusion mechanism composes three types of convolution filters: $\mathbf{f_1} \in \mathcal{R}^{c \times 1}$, $\mathbf{f_2} \in \mathcal{R}^{c \times 2}$, and $\mathbf{f_3} \in \mathcal{R}^{c \times 3}$. $\mathbf{f_1}$ can capture the underlying information for each feature vector in $\mathbf{z_{i,j}}$; $\mathbf{f_2}$ captures the correlation information between $\mathbf{p_{u}^{i}}$ and $\mathbf{t_{i,j}}$; $\mathbf{f_3}$ serves to learn the overall interaction across all the three feature vectors in $\mathbf{z_{u,i}}$.
Then, we apply max-pooling operations after each convolutional layer.
The operations obtain the comprehensive features from both  user and item.
We denote the features as $\mathbf{m_{i,j}} \in \mathcal{R}^{K \times 1}$, where $K$ is the number of all filters. 
Finally, we stack a fully connected layer to transform the feature dimension of $\mathbf{m_{i,j}}$ as follows:
\begin{equation}\label{cut d}
    \mathbf{h_{i,j}} = \mathbf{m^\top_{i,j}} \mathbf{W_{1}} + \mathbf{b_{1}} \ 
\end{equation}
where $\mathbf{W_{1}} \in \mathcal{R}^{c \times K}$ is the weight matrix, $\mathbf{b_{1}} \in \mathcal{R}^{c \times 1}$ is the bias.
The $\mathbf{h_{i,j}} \in \mathcal{R}^{c \times 1}$ is the final feature vector for the given user-item pair, which incorporates the score feature, the interactive feature, and their hidden correlation.

\subsection{Skip Connection Module}
While the Gumbel-based module can balance skewed features, an existing research~\cite{he2016deep} shows that as network depth grows, neural models can suffer gradient degradation challenges and can not efficiently utilize enough input features, such as $\alpha$ and $\beta$.
We deploy a skip connection module to re-connect rating frequencies $\mathbf{freq_{u}^{i}}$ with the feature $\mathbf{h_{i,j}}$ to alleviate the vanishing gradient of the score feature.
Skip connection is a key module in residual network~\cite{he2016deep} allowing networks to feed one layer outputs to several next layers without following order and reduce gradient vanishing issue during training periods.
We adapt the skip connection to capture patterns of user ratings by incorporating the fused features and user rating pattern (frequency distribution) for final rating predictions.
We do not add item rating pattern because of its little improvement in our experiments.
Our final prediction function is as follows,
\begin{equation}\label{pre formula}
     \hat{r}_{i,j} = \sum_{l=1}^{c} f(freq_{u,l}^{i}) h_{i,j,l}\  
\end{equation}
, where $f(\cdot)$ is a fully connected layer and the output has the same dimension as the input.

Rating prediction has two primary recommendation tasks, regression and ranking. 
To evaluate our proposed method, we deploy two types of optimization objectives in Recommender System, rating regression and top-N ranking.
1) For the regression optimization objective, we utilize the square loss function to train our model as follows:
\begin{equation}
\mathcal{L} = \sum_{i,j \in \mathbf{\Omega}} (\hat{r}_{i,j} - r_{i,j})^2
\end{equation}
, where $\mathbf{\Omega}$ denotes the set of instances (i.e., users and items) for training, and $r_{i,j}$ is the ground truth rating assigned by the user $i$ to the item $j$.
The regression task goal is to select an item list by predicting top rating scores.
2) For the top-N recommendation evaluation, we define the objective function as binary
cross-entropy loss function:
\begin{equation}
    \mathcal{L}=-\sum_{u, i} \mathbf{r}_{u i} \log \hat{\mathbf{r}}_{u i}+\left(1-\mathbf{r}_{u i}\right) \log \left(1-\hat{\mathbf{r}}_{u i}\right)
\end{equation}
The ranking task goal is to select an item list that users are likely to buy.
We choose the rating records as the ranking optimization targets.
If a user $u$ rates an item $i$, then we can obtain $\mathbf{r}_{u i}=1$, otherwise $\mathbf{r}_{u i}=0$.

\section{Experiments}

In this section, we present experimental settings of our approach and baselines.
We keep the review length and other review numbers covering 0.85 percent of users and items: we pad the short reviews and truncate long reviews to the same length.
Our experiments randomly split each dataset into training set (80\%), validation set (10\%), and test set (10\%).
We compare our method with rating- and review-based baselines with conventional and recent class-balancing approaches.


\subsection{Baselines}
We select recent methods in both rating- and review-based models in the rating prediction task as baselines.
The implementation follows the default settings from baseline studies.
Note that in our GVN framework, we can also deploy these baselines as feature extractors on the input data (i.e., the gray rectangle in Figure \ref{fig:overview}).

\noindent (1) Rating-based methods:
\begin{itemize}
    \item \textbf{PMF}~\cite{mnih2008probabilistic} is a standard matrix factorization to model users and items from rating matrix. We set the length of users' and items' embedding feature as 50 in this baseline.
    
    \item \textbf{NeuMF}~\cite{he2017neuralcf} characterizes an interactive feature of user-item pairs by concatenating the results learned from both matrix
    factorization and multi-layer perceptron (MLP). The model follows original work with 10 MLP layers with 32 factor neurons per layer.
\end{itemize}
(2) Review-based methods:
\begin{itemize}
    \item \textbf{DeepCoNN}~\cite{zheng2017joint} obtains the convolutional representations of user's and item's reviews and passes the concatenated representations into a factorization machine (FM) model. The convolutional network set kernel number as 20, kernel size as 3, and a max pooling layer.
    
    \item \textbf{ANR}~\cite{chin2018anr} captures aspect-based features of users and items by a co-attention network over different aspects. ANR contains 5 aspects and sets the dimensionality of each aspect as 10.
    
    \item \textbf{NARRE}~\cite{chen2018neural} utilizes two neural networks (hidden size as 32) with an attention mechanism to weigh useful reviews. NARRE learns the interactive features of users, items, and reviews. We set vector dimensions of user and items as 32.
    
    \item \textbf{NRPA}~\cite{liu2019nrpa} uses a hierarchical attention network to learn personalized features for users and items. NRPA uses the user and item embeddings with the same dimension as NARRE.
    
    \item \textbf{TDAR}~\cite{yu2020semi} proposes a Text Memory Network to characterize the users' and items' features from reviews. The Text Memory Network includes three networks (dimension as 128), users, items, and tokens. The word network encodes reviews by a pre-trained embedding and projects into the same dimension as users and items. 
\end{itemize}

(3) Imbalanced label methods:
\begin{itemize}
    \item  \textbf{RERANK}~\cite{wei2021towards} achieves the state-of-the-art performance in imbalance rating prediction by training head and tail label samples separately on two models.
    Our implementation set the 1-,2-, and 3-ratings as tail label samples and 4-, and 5- ratings as head label samples.
\end{itemize}

\subsection{Experimental Setting}
We set all the parameters in the baseline methods the same as mentioned in their original papers.
All the models utilize 300-D word embedding trained on Wikipedia by using GloVe~\cite{pennington2014glove} and Adam~\cite{diederik2016adam} optimizer updates the trainable parameters.
We set GVN and the baselines with a learning rate among $[0.005,0.01,0.03]$, the learning rate to decay by $60\%$ every epoch, and the batch size as $64$. 
We train all the models for a maximum of $30$ epochs and report the result on the corresponding test set when the validation set achieves the best result.
For the baselines not for imbalanced ratings, we apply conventional ways to re-weight unbalanced ratings.
Our experiment repeats five times, and all records are statistically significant at 0.05 level by t-test~\cite{joan1987Statistical}.

\paragraph{Evaluation Metrics}
Error- and ranking-based metrics are two common evaluations in Recommender System~\cite{chen2018neural, zhao2020improving}.
We evaluate our model by both metrics to prove the generality of GVN.
In error-based metrics, we use Mean Absolute Error to evaluate the predicting results of the rare data.
The MAE is robust for imbalanced data because the measurement is robust to outliers~\cite{liu2019daml}. 
In ranking-based metrics, we employ F1 score, Hit Ratio (HR)~\cite{chen2018neural}, Normalized Discounted Cumulative Gain and (NDCG)~\cite{jarvelin2020cumulated} to evaluate the items recommendation list.
$HR@k$ measures the proportion of cases that the desired item is among the top-k items in all test cases.
$NDCG@k$ is a position-aware metric which assigns larger weights on higher ranks. 
Following the previous work \cite{kang2018attentive}, for each test user, 
GVN predicts the score of each negative item and positive item in the test set and gives a recommendation list according to ranking.


\begin{figure*}[htp]
\centering
\includegraphics[width=0.87\textwidth]{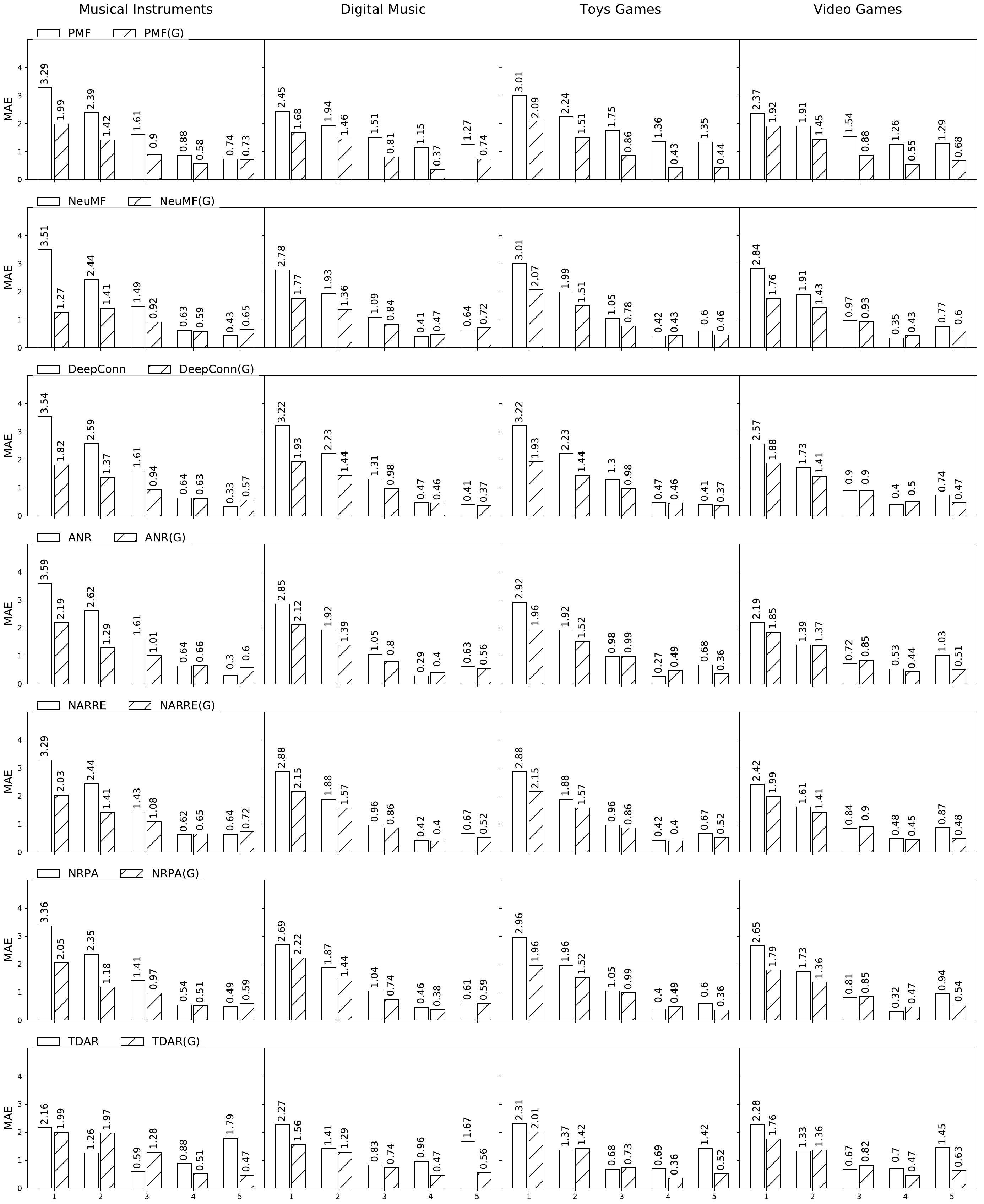}
\caption{Prediction error (MAE) for each rating. The numbers on the horizontal axis represent different ratings}
\label{fig:all_score_loss}
\end{figure*}

\subsection{Performance Evaluation}
\begin{table*} 
	\centering
 \caption{Performance comparison (Top-5) on five benchmark datasets.}
		\resizebox{1\textwidth}{!}
	{\begin{tabular}{l |c c c| c c c| c c c| c c c | c c c}
	\toprule
		~ & \multicolumn{3}{c}{Musical Instrument} & \multicolumn{3}{c}{Digital Music} & \multicolumn{3}{c}{Toys Game} & \multicolumn{3}{c}{Video Game} &\multicolumn{3}{c}{Yelp}\\
		\midrule
		Top@5 &F1 &HR &NDCG &F1 &HR &NDCG &F1 &HR &NDCG &F1 &HR &NDCG &F1 &HR &NDCG\\
		\hline
		PMF &0.0265 &0.0903 &0.0334 &0.0633& 0.2437&0.0745 &0.0210 &0.0710 &0.0359 &0.0436 &0.1812 &0.0595 &0.1515 &0.6900 &0.2400\\
		PMF(G) &\bf 0.1629 &\bf 0.4811 &\bf 0.2511 &\bf 0.1696 &\bf 0.5254 &\bf 0.2212 &\bf 0.1405 &\bf 0.4343 &\bf 0.1907 &\bf 0.2190 &\bf 0.5958 &\bf 0.2703 &\bf 0.1888 &\bf 0.7540 &\bf 0.2850\\
		\hline
		NeuMF &0.0493 &0.1686 &0.0617 &0.0819 &0.3014 &0.1012 &0.0646 &0.2259 &0.0763 &0.0319 &0.1220 &0.0355 &0.2114 &0.7840 &0.3298\\
		NeuMF(G) &\bf 0.1768 &\bf 0.5119 &\bf 0.2738 &\bf 0.1552 &\bf 0.5015 &\bf 0.1998 &\bf 0.1049 &\bf 0.3420 &\bf 0.1451 &\bf 0.2197 &\bf 0.5828 &\bf 0.2727 &\bf 0.2181 &\bf 0.8080 &\bf 0.3320\\
		\hline 
		DeepCoNN &0.0383 &0.1329 &0.0441 &0.1014 &0.3557 &0.1318& 0.1211 &0.3049 &0.1334 &0.1183 &0.3732 &0.1364 &0.2014 &0.7460 &0.3154\\
		DeepCoNN(G) &\bf 0.1654 &\bf 0.4831 &\bf 0.2635 &\bf 0.1669 &\bf 0.4850 &\bf 0.2050 &\bf 0.1291 &\bf 0.3060 &\bf 0.1441 &\bf 0.1232 &\bf 0.3820 &\bf 0.1387 &\bf 0.2150 &\bf 0.7940 &\bf 0.3297\\
		\hline
        NARRE &0.0880 &0.2788 &0.1144 &0.1312 &0.4339 &0.1671 &0.1221 &\bf 0.3520 &0.1197 &0.1365 &0.4266 &0.1590 &0.2099 &0.7880 &\bf 0.3302\\
        NARRE(G) &\bf 0.1669 &\bf 0.5000  &\bf 0.2749 &\bf 0.1595 &\bf 0.5104 &\bf 0.2308 &\bf 0.1252 &0.3280 &\bf 0.1532 &\bf 0.1404 &\bf 0.4384 &\bf 0.1661 &\bf 0.2144 &\bf 0.8000 &0.3233 \\
        \hline
        NRPA &0.0493 &0.1636 &0.0547 &0.0135 &0.0256 &0.0088 &0.0416 &\bf 0.1549 &\bf 0.0547 &0.0537 &0.1978 &0.0582 &0.2069 &\bf 0.7960 &0.3271\\
        NRPA(G) &\bf 0.1607 &\bf 0.4702 &\bf 0.2453  &\bf 0.0418 &\bf 0.0590 &\bf 0.0150 &\bf 0.0467 &0.1260 & 0.0532 &\bf 0.2083 &\bf 0.5664 &\bf 0.2528 &\bf 0.2169 &0.7840 &\bf 0.3314\\
        \hline
        TDAR &0.1171 &0.3629 &0.1500 &0.1151 &0.4049 &0.1446 &0.1491 &0.4377 &0.1899 &0.1410 &0.4428 &0.1726 &0.1038 &0.5168 &0.1475\\
        TDAR(G) &\bf 0.1672	&\bf 0.4970	&\bf 0.2500 &\bf 0.1498	&\bf 0.4933	&\bf 0.2008 &\bf 0.1534	&\bf 0.4465	&\bf 0.2063 &\bf 0.1837	&\bf 0.5302	&\bf 0.2247 &\bf 0.1107 &\bf 0.5603 &\bf 0.1631\\
		\hline
		\bottomrule
	  \end{tabular}}
	  \label{tb:comparison2}
\end{table*}

In this section, we evaluate the overall performance by using error-based and ranking-based metrics.
As shown in Table \ref{tb:comparison} and \ref{tb:comparison2}, we report the comparison of baselines and the methods as base models within GVN.
For example, NARRE(G) denotes a model that utilizes GVN framework to train NARRE. 
As shown in Table~\ref{tb:comparison}, the overall prediction performance of our method significantly outperforms several baselines. 
Our method achieves a substantial improvement of about 0.46\% to 57.82\% on all four datasets.

The most significant improvement is on TDAR, and all datasets gain improvement over 50\%. 
We infer the reason behind this is that TDAR is for classification tasks and mainly focuses on the Top-N performance, so it cannot solve the rating prediction problem in regression tasks. 
Besides, GVN also gains a remarkable improvement on TDAR in ranking performance (shown in Table~\ref{tb:comparison}). 
Since TDAR is a state-of-the-art model for ranking performance, we can observe that it surpasses other baselines on all datasets with respect to F1 score, HR, and NDCG. 
However, after adding GVN, other traditional models obtain a comparable performance with TDAR. 

\begin{table}[htp]
\centering
 \caption{Performance comparison (MAE). The best results are in bold. $\triangle \%$ denotes percentages of relative improvement. }  
	\resizebox{0.45\textwidth}{!}
	{\begin{tabular}{l c c c c c c c c c c}
	\toprule
		\multirow{2}*{Method} &Musical &Toys &Digital &Video &Yelp\\
		~ &Instrument & Game &Music  &Game\\
		\midrule
		Rating-based&&&& \\
		\hline
		PMF &0.744	&0.788	&0.895	&1.421 &0.738	\\
		PMF(G) &\bf 0.639 &\bf 0.566	&\bf 0.662	&\bf 0.799 &\bf 0.725\\
		$\triangle\%$ &14.11 &28.17	&26.03	&43.82 &1.76\\
		\hline
		NeuMF  &0.655 &0.723	&0.778	&0.902 &0.798\\
		NeuMF(G) &\bf 0.652	&\bf 0.552	&\bf 0.689	&\bf 0.720 &\bf 0.780\\ 
		$\triangle\%$ &0.460	&23.65	&11.44	&20.22 &2.26\\
        \midrule
        Review-based&&&& \\
        \midrule
        DeepCoNN &\bf 0.636	&0.664	&0.769	&0.860 &0.768\\
	    DeepCoNN(G)	&0.638	&\bf 0.538	&\bf 0.532	&\bf 0.668 &\bf 0.728\\
	    $\triangle\%$ &-0.31 &18.98	&30.82	&22.27 &5.21\\
	    \hline
	    ANR &0.623  &0.657 &0.766 &0.987 &0.783\\
	    ANR(G) &\bf0.611 &\bf0.510 &\bf0.675 &\bf0.664 &\bf0.760\\
	    $\triangle\%$ &1.93 &22.37 &11.88 &32.73 &2.94 \\
	    \hline
	    NARRE	&0.771	&0.710	&0.802	&0.926  &0.860 \\
	    NARRE(G)	&\bf 0.665	&\bf 0.534	&\bf 0.669	&\bf 0.670 &\bf 0.820\\
	    $\triangle\%$ &13.75 &24.79	&16.58 &27.64 &4.65\\
	    \hline
	    NRPA	&0.684	&0.692	&0.806	&0.949 &0.881\\
	    NRPA(G) &\bf 0.616	&\bf 0.570	&\bf 0.543	&\bf 0.683 &\bf 0.726\\
	    $\triangle\%$ &11.04 &21.39	 &32.63 &28.05 &17.59\\
	    \hline
	    TDAR  &1.504 &1.375 &1.179 &1.248 &1.195\\
	    TDAR(G) &\bf 0.634 &\bf 0.649 &\bf 0.570 &\bf 0.727 &\bf 1.052\\
	    $\triangle\%$ &57.82 &52.84 &51.65 &54.3 &11.96\\
		\bottomrule
	  \end{tabular}}
	  \label{tb:comparison}
\end{table}

 We find that the loss of the rating prediction task positively correlates to the percentage of low ratings.
Because the extra capability to distinguish the low and high ratings learned from the slight increase number of low rating samples cannot offset the negative influence that models need to predict more unfamiliar samples.
Therefore, the inferior ability to predict low rating samples is gradually exposed with the increase of low rating samples. 
Thus, we find that GVN improves baselines on the Video Game to a noticeable extent, in which low rating samples account for the greatest ratio compared to the other four datasets.
When the imbalance between low and high rating samples is narrowed, the model's performance will be crippled and then increased.
As shown in the Table~\ref{tb:data}), 4-score accounts for the most considerable amount, and the ratio of 3-score is the highest among other datasets, which significantly increases the prediction accuracy on rare samples. 
We also find Gumbel distribution achieves better performance than other distributions in imbalanced data.
For example, the Gumbel distribution outperforms other distributions (e.g., Poisson and Normal). While both Weibull and Gumbel distributions belong to extreme value distributions (EVD), the Gumbel distribution shows its supremacy over the other EVDs.

\subsection{Prediction for Rare Ratings}

In this section, we show the detailed prediction error of each ratings to verify the effectiveness of GVN for a biased prediction problem with long-tailed ratings. 
Figure~\ref{fig:all_score_loss} shows MAE scores on each individual rating class as histograms.
Lower histograms means lower MAE scores and better performance.

We experimentally prove that after applying GVN framework, all baselines achieve a significant improvement on rare ratings.
Additionally, our improvement on rare data (tail classes) does not sacrifice the performance of common data (head classes).
We notice that TDAR outperforms other baselines on rare data but gets poorer overall performance than other baselines. It is because TDAR fails to predict frequent samples. 
For example, TDAR has about 1.5 MAE loss on 5-score samples on Toys Games.
Also, the samples with 1 and 2 scores have the most significant prediction error on Toys Game, which is because the low rating samples account for the low ratio (shown as Fig.\ref{fig:samplex}. 1-, 2-, 3-, 4-, 5-score account for $2.8\%$, $3.8\%$, $9.8\%$, $22\%$ and $61\%$ respectively). 
We infer that our framework can successfully leverage user preferences in the imbalanced settings.
Notice that our framework also achieves better performance on overall performance across the five datasets.
This experimentally highlights that our end-to-end framework can not only solve the biased prediction challenge, but also generally improve overall performance in the rating prediction task.

In some situations, the loss of baselines is lower than GVN. 
For example, the loss of ANR prediction on 4-score is lower than ANR(G).
The reason is that some models nearly indistinguishably predict the 4-score for all samples.
If we hope to improve overall performance, samples with 4-score inevitably suffer from the increase of loss.
The goal of the recommender system is not to predict a precise score on a specific rating sample but to distinguish the difference of user' preference. 
Therefore, we should concentrate on both the overall rating performance and ranking metrics.


\begin{table}[ht]
\centering
\caption{Imbalance Methods Performance Comparisons.}
\label{table4}
\resizebox{0.46\textwidth}{!}
{
    \begin{tabular}{c||cc||cccc}
    \multirow{2}{*}{Data} & \multirow{2}{*}{Method} & \multirow{2}{*}{Metrics} & \multicolumn{4}{c}{Base Model} \\
     &  &  & PMF & NeuMF & NARRE & NRPA \\\hline\hline
    \multirow{8}{*}{Amazon} & \multirow{4}{*}{RERANK} & MAE & 0.937 & 0.847 & 1.408 & 1.201 \\
     &  & F1 & 0.070 & 0.075 & 0.062 & 0.067 \\
     &  & HR & 0.232 & 0.248 & 0.202 & 0.228 \\
     &  & NDCG & 0.900 & 0.101 & 0.082 & 0.085 \\\cline{2-7}
     & \multirow{4}{*}{GVN} & MAE & \bf 0.639 & \bf 0.652 & \bf 0.665 & \bf 0.616 \\
     &  & F1 & \bf0.163 & \bf0.177 & \bf0.167 & \bf0.161 \\
     &  & HR & \bf0.481 & \bf0.512 & \bf0.500 & \bf 0.470 \\
     &  & NDCG & \bf0.251 & \bf0.274 & \bf0.275 & \bf 0.245 \\
    \hline\hline
    \multirow{8}{*}{Yelp} & \multirow{4}{*}{RERANK} & MAE & 0.882 & 1.428 & 0.842 & 0.895 \\
     &  & F1 & 0.118 & 0.107 & 0.132 & 0.107 \\
     &  & HR & 0.574 & 0.558 & 0.622 & 0.536 \\
     &  & NDCG & 0.174 & 0.148 & 0.207 & 0.161 \\\cline{2-7}
     & \multirow{4}{*}{GVN} & MAE & \bf 0.725 & \bf 0.780 & \bf 0.820 & \bf 0.726 \\
     &  & F1 & \bf0.189 & \bf0.218 & \bf0.214 & \bf0.217 \\
     &  & HR & \bf 0.754 & \bf0.808 & \bf0.800 & \bf0.784 \\
     &  & NDCG & \bf0.285 & \bf0.332 & \bf0.323 & \bf0.331
    \end{tabular}
}
\end{table}

\subsection{Imbalance Methods Comparison}
While we compare conventional imbalance approach using re-weighting in the previous evaluation sections, we compare GVN with the state-of-the-art imbalanced label methods in this section. 
We present results in Table~\ref{table4}.
We select four base models and conduct this experiment on two datasets. 
The result shows that the model with ReRank is consistently lower than the original models on the Amazon dataset, and rare rating samples account for 11.57 percent of overall samples.
Therefore, the model only learns from long-tail label samples overfit to small samples, which deteriorates the overall performance.
NARRE-ReRank obtains marginal improvement on yelp by comparing the NARRE model. Overall, GVN improves significantly than ReRank on all base models and datasets.

\subsection{Ablation Study of GVN} 
\begin{table}[ht]
	\centering
	\caption{Ablation studies on different components of the proposed GVN. Lower scores are better.}
	\resizebox{0.45\textwidth}{!}
	{
	\begin{tabular}{l c c c c c }
	\toprule
	\multirow{2}*{Method} &Musical &Office &Toys &Grocery \\
    ~ &Instrument&Product &Game &Food  \\
    \midrule
	PMF(G) & \bf 0.639 &0.584 &0.566	& \bf 0.678\\
	-G &0.694	&0.703	&0.630	&0.744\\
	-F &0.790	&0.631	&\bf 0.554	&0.754\\
	-S &0.640	&\bf 0.581	&0.594	&0.709\\
	\hline
	NeuMF(G) & \bf 0.652	&0.622	& \bf0.552	& 0.750\\
	-G &0.734	&0.658	&0.604	&0.770\\
	-F &0.723	&0.676	&0.592	&\bf 0.731 \\
	-S &0.700	&\bf 0.621	&0.598	&0.767\\
	\hline
    DeepCoNN(G) &0.638	&\bf 0.600 &\bf0.538	&\bf 0.665\\
	-G &0.711	&0.644	&0.576	&0.719\\
	-F &0.677	&0.637	&0.539	&0.737\\
	 -S &0.637	&0.607	&0.557	&0.676\\
    \hline
    NRPA(G) &\bf 0.616	&0.609	&\bf0.570	&\bf0.708\\
    -G &0.660	&0.701	&0.605	&0.779\\
    -F &0.766	&0.664	&0.648	&0.771\\
    -S &0.679	&\bf0.580	&0.595	&0.732\\
    \hline
    NARRE(G) &0.665	&\bf 0.592	&\bf 0.534	&\bf 0.669\\
    -G &0.790	&0.680	&0.586	&0.726\\
    -F &0.727	&0.738	&0.572	&0.726\\
    -S &\bf 0.644	&0.599	&0.569	&0.741\\
    \hline
    ANR(G) &\bf 0.611 &\bf 0.604 &\bf 0.510 &0.754\\
    -G &0.733 &0.676 &0.563 &\bf 0.706\\
    -F &0.649 &0.650 &0.550 &0.711\\
    -S &0.638 &0.627 &0.545 &0.723\\
	\bottomrule
	\end{tabular}}
	\label{tb:comparison detailX}
\end{table}
We conduct ablation studies to quantify the influence of three main modules in GVN framework, including the Gumbel-based feature learning module, the feature fusion layer, and the skip connection prediction module.

The comparison models are defined as follows: 1) \textbf{-G}: This model removes the Gumbel-based feature learning module, which means the original rating frequencies directly feed into the following fusion layer as the score features; 2) \textbf{-F}: A model with the infusion layer. We use the average of user score feature, item score feature, and interactive feature as the final representation, instead of using the original feature fusion layer (i.e., multi-scale convolutional neural network) to summarize the features mentioned above;
3) \textbf{-S}: 
A model that removes the skip connection prediction module, and the prediction result is ${\hat{r}_{i,j}}=\sum_{l=1}^{c}h_{i,j,l}\ $ instead of E.q.\ref{pre formula}.
The comparison results in Table~\ref{tb:comparison detailX} show that the methods with GVN outperform most of the ablation models.
The performance of models that removing the Gumbel-based feature learning module drops significantly compared, which shows the effectiveness of the proposed Gumbel-based module.
Moreover, comparing with all the original models without applying any components of GVN, the \textbf{-G} models marginally improve their overall performance, which demonstrates the necessity of utilizing the proposed multi-scale feature fusion layer, the skip connection prediction module.

Several \textbf{-S} and \textbf{-F} methods can obtain slightly better performance on some datasets than those with complete GVN architecture (i.e., \textbf{(G)} methods).
For example, NARRE has modest performance improvement after removing the skip connection prediction module on the Musical Instrument dataset. PMF achieves the best performance on the Toys Game dataset without utilizing the feature fusion layer. 
However, in most cases,  using the complete GVN architecture can obtain the best results, which further demonstrates the necessity of jointly using the proposed three components for rating prediction.
We conclude that the Gumbel-based module plays a major role in improving the overall performance of the rating prediction models. 
Moreover, the multi-scale feature fusion layer and the skip connection prediction module are auxiliary components that further improve the proposed model's performance. 
While we develop the Gumbel-based modules to alleviate the biased prediction problem, the mechanism will apply to other prediction tasks to improve their overall performance.
\subsection{Different Distributions Comparisons}
\begin{table}[ht!]
	\centering
	\caption{The impact of using several different types of distributions on GVN. Note that (G) is Gumbel distribution; (P) is Poisson distribution; (N) represents Normal distribution; (E) represents Exponential distribution; (W) is Weibull distribution and (F) is Fréchet distribution. Lower scores are better.}
\resizebox{0.45\textwidth}{!}{
	\begin{tabular}{l c c c c c }
	\toprule
	\multirow{2}*{Method} &Musical &Office &Toys &Grocery \\
    ~ &Instrument&Product &Game &Food  \\
    \midrule
	PMF(G) & \bf 0.639 &\bf 0.584 &0.566	& \bf 0.678\\
	PMF(P) &0.692 &0.700 &0.594 &0.748\\
	PMF(N) &0.683 &0.618 &0.560 &0.684\\
    PMF(E) &0.740 &0.684 &0.587 &0.753\\
    PMF(W) &0.696 &0.590 &0.578 &0.697\\
    PMF(F) &0.686 &0.607 &\bf0.551 &0.703\\
	\hline
	NeuMF &0.655 &0.669	&0.723	&0.793\\
	NeuMF(G) & 0.652	&\bf0.622	& \bf0.552	& \bf0.750\\
	NeuMF(P) &0.679 &0.735 &0.659 &0.824\\
	NeuMF(N) &0.735 &0.704 &0.721 &0.778\\
	NeuMF(E) &\bf0.651 &0.722 &0.719 &0.828\\
	NeuMF(W) &0.715 &0.722 &0.713 &0.834\\
	NeuMF(F) &0.744 &0.740 &0.735 &0.753\\
	\hline
    DeepCoNN(G) &0.638	&\bf 0.600 &0.538	&\bf 0.665\\
	DeepCoNN(P) &0.628 &0.648 &0.562 &0.749\\
	DeepCoNN(N) &\bf0.627 &0.609 &\bf0.521 &0.666\\
	DeepCoNN(E) &0.702 &0.650 &0.564 &0.688\\
	DeepCoNN(W) &0.660 &0.624 &0.547 &0.694\\
	DeepCoNN(F) &0.653 &0.618 &0.552 &0.712\\
    \hline
    NRPA(G) &\bf 0.616	&\bf0.609	&0.570	&0.708\\
    NRPA(P) &0.691 &0.675 &0.646 &0.781\\
	NRPA(N) &0.655 &0.629 &\bf0.550 &\bf0.691\\
	NRPA(E) &0.673 &0.713 &0.638 &0.744\\
	NRPA(W) &0.677 &0.668 &0.547 &0.699\\
	NRPA(F) &0.746 &0.663 &0.569 &0.698\\
    \hline
    NARRE(G) &\bf0.665	&0.592	&\bf 0.534	&\bf 0.669\\
    NARRE(P) &0.666 &0.689 &0.600 &0.710\\
	NARRE(N) &0.678 &0.675 &0.548 &0.680\\
	NARRE(E) &0.740 &0.684 &0.587 &0.753\\
	NARRE(W) &0.696 &\bf0.590 &0.578 &0.686\\
	NARRE(F) &0.686 &0.607 &0.551 &0.672\\
    \hline
    ANR(G) &\bf 0.611 &0.604 &\bf 0.510 &0.754\\
    ANR(P) &0.629 &0.655 &0.553 &0.698\\
	ANR(N) &0.631 &0.600 &0.523 &0.665\\
	ANR(E) &0.636 &0.653 &0.626 &0.742\\
	ANR(W) &0.628 &\bf0.572 &0.541 &\bf0.678\\
	ANR(F) &0.735 &0.624 &0.517 &0.704\\
	\bottomrule
	\end{tabular}}
	\label{tb:comparison detail3}
\end{table}
In this section, we conduct performance comparison across distributions in rating prediction. 
We follow the same experimental settings with our GVN framework, and replace the Gumbel module by the other distribution functions.
We include three general distributions and two extreme value distributions (EVDs). 
The three distributions are Poisson distribution (P), Normal distribution (N) and Exponential distribution (E), and the EVDs are Weibull distribution (W), Fréchet distribution (F). 
We demote our Gumbel distribution as (G).
The comparison serves to verify if the Gumbel distribution has advantages in rating prediction task and biased prediction challenge.

We evaluate the models by mean average error (MAE) and present the results in Table~\ref{tb:comparison detail3}.
Gumbel-based models generally outperform other distributions across the four datasets by a large margin.
The GVN framework achieves the best performance in 75\% of all tasks (18 tasks out of 24 tasks).
Those demonstrate the effectiveness our proposed framework and suggest that the proposed modules can augment rating prediction models performance in understand user preferences.
Additionally, we can find that the proposed framework with different distributions can generally improve performance of baseline models.
This highlights that our proposed framework can significantly augment overall performance of baselines.
We can also infer that for imbalanced datasets in rating prediction task, deploying distributions to balance skewed ratings is critical.
Moreover, the extreme value distributions (Gumbel, Weibull, and Fréchet) achieve better performance than the other distributions in major tasks.
This indicates that EVDs are more effective than regular distributions (e.g., Gaussian distribution) and Gumbel distribution has more advantages than the other two EVDs for imbalanced datasets in this rating prediction task. 

\section{Related work}
\textbf{Rating- and Review-based models} are general methods in the rating prediction task.
Common rating-based methods include Factorization Machine (FM)~\cite{rendle2010factorization}, Probabilistic Matrix Factorization (PMF)~\cite{salakhutdinov2007probabilistic}, and Latent Factor Model (LFM)~\cite{koren2009matrix}. 
MF and PMF decompose the observed rating matrix into a product of two low-rank latent feature matrices.
Neural Matrix Factorization (NeuMF)~\cite{he2017neuralcf} combines with representations obtained by multi-layer perceptron networks.
LFM introduces the user, item, and global variables to correct errors caused by personal score preference.
However, the rating matrix is usually natural sparsity, which can make it difficult for the rating-based models to learn the accurate latent features of users and items~\cite{chen2018neural}.
To tackle the limitations of rating-based methods, a number of studies derive representations of users and items from user textual reviews~\cite{wang2021Leveraging, yu2020semi, li2019capsule}.
For example, \citeauthor{liu2019nrpa} propose a model that encodes user reviews with neural transformers \cite{vaswani2017attention} and uses FM to decompose interactive vector.
However, the rating- and review-based methods do not explicitly consider rare ratings in imbalanced datasets leading to a biased prediction. 
We experimentally demonstrate the biased prediction of the methods in Table~\ref{tb:sample} that the three methods predict towards head classes of user ratings.
In this work, we proposed a Gumbel-based framework to handle the biased prediction for imbalanced rating classes.

\textbf{Imbalanced Class} exists when sample sizes of some classes are significantly larger than other classes. 
The imbalanced dataset is a leading factor for the biased prediction that models will predict towards head classes than tail classes. 
Data resampling heuristics and cost-sensitive learning are two common strategies for addressing the biased prediction problem.
Data resampling heuristics achieve a balanced data distribution by over-sampling or under-sampling the original imbalanced data~\cite{estabrooks2004multiple, drummond2003c4,yuan2020on,mahajan2018exploring}.
Cost-sensitive learning aims at assigning weights of loss to rating classes~\cite{drummond2003c4,kini2021advances,lin2020focal}.
However, these methods depend on domain knowledge and require human effort to set up relevant hyperparameters (i.e., class weights).
Some recent works proposed to train each label on different models~\cite{wei2021towards,babbar2017dismec,khan2019striking,cui2019class}. However, these methods increase the memory overhead and fail to capture the correlation between different labels. 

In this study, we propose an end-to-end neural framework based on Gumbel distribution that coherently model imbalanced classes in a data-driven approach.
Gumbel-Softmax trick~\cite{jang2017categorical} is a close work that combines reparameterization technique and softmax function to replace the non-differentiable sampling process.
A number of studies utilize the Gumbel-Softmax to improve quality of discrete data generation, such as text generation~\cite{zhi2020catgan}, machine translation~\cite{gu2018neural} and image generation~\cite{jang2017categorical}.
Unlike the Gumbel-Softmax trick, our study has two major differences.
First, our proposed GVN is a deterministic process that uses the Minimum Gumbel PDF as an activation function to model imbalanced ratings. 
Second, GVN deploys Gumbel distribution on rating prediction task, which has not been explored in the previous studies.

\section{Conclusion}

In this paper, we propose a Gumbel-based Rating Prediction (GVN) framework to address the biased prediction problem in the rating prediction task. 
Our framework utilizes the Gumbel distribution to define both head and tail rating classes.
Extensive experiments can demonstrate that our framework can effectively solve the biased prediction problem and achieve the state-of-the-art performance across all benchmarks.
We also demonstrate that Gumbel distribution is more effective to model imbalanced ratings than the other types of distributions.
To our best knowledge, GVN is the first work to address the biased prediction problem using the Gumbel distribution in the rating prediction task.
Moreover, while we focus on rating prediction task, our proposed method provides a flexible solution for the imbalanced classes problem in other fields.

\section{Acknowledgement}
The authors thank reviewers for their valuable comments. 

\bibliographystyle{ACM-Reference-Format}
\bibliography{sample-base}

\end{document}